\documentclass[twocolumn]{aastex62}
\usepackage{apjfonts}
\usepackage{lineno}

\received{2018 September 18}
\revised{2018 October 8}
\accepted{2018 October 8}
\submitjournal{ApJL}

\shorttitle{{\it NuSTAR} Detection of Nonthermal Bremsstrahlung from W49B}
\shortauthors{Tanaka et al.}

\begin{document}

\title{{\it NuSTAR} Detection of Nonthermal Bremsstrahlung from the Supernova Remnant W49B}

\author{Takaaki Tanaka}
\correspondingauthor{Takaaki Tanaka}
\affil{Department of Physics, Kyoto University, Kitashirakawa Oiwake-cho, Sakyo, Kyoto 606-8502, Japan}
\email{ttanaka@cr.scphys.kyoto-u.ac.jp}

\author{Hiroya Yamaguchi}
\affil{Institute of Space and Astronautical Science, JAXA, 3-1-1 Yoshinodai, Sagamihara, Chuo, Kanagawa 252-5210, Japan}

\author{Daniel R. Wik}
\affil{Department of Physics \& Astronomy, The University of Utah, 115 South 1400 East, Salt Lake City, UT 84112, USA}

\author{Felix A. Aharonian}
\affil{Dublin Institute for Advanced Studies, 31 Fitzwilliam Place, Dublin 2, Ireland}
\affil{Max-Planck-Institut f\"ur Kernphysik, P.O. Box 103980, D-69029 Heidelberg, Germany}
\affil{MEPHI, Kashirskoe shosse 31, 115409 Moscow, Russia}

\author{Aya Bamba}
\affil{Department of Physics, The University of Tokyo, 7-3-1 Hongo, Bunkyo, Tokyo 113-0033, Japan}
\affil{Research Center for the Early Universe, The University of Tokyo, 7-3-1 Hongo, Bunkyo, Tokyo 113-0033, Japan}

\author{Daniel Castro}
\affil{Harvard-Smithsonian Center for Astrophysics, 60 Garden Street, Cambridge, MA 02138, USA}

\author{Adam R. Foster}
\affil{Harvard-Smithsonian Center for Astrophysics, 60 Garden Street, Cambridge, MA 02138, USA}

\author{Robert Petre}
\affil{NASA Goddard Space Flight Center, Code 662, Greenbelt, MD 20771, USA}

\author{Jeonghee Rho}
\affil{SETI Institute, 189 N. Bernardo Ave., Mountain View, CA 94043, USA}
\affil{SOFIA Science Center, NASA Ames Research Center, MS 232, Moffett Field, CA 94035, USA}

\author{Randall K. Smith}
\affil{Harvard-Smithsonian Center for Astrophysics, 60 Garden Street, Cambridge, MA 02138, USA}

\author{Hiroyuki Uchida}
\affil{Department of Physics, Kyoto University, Kitashirakawa Oiwake-cho, Sakyo, Kyoto 606-8502, Japan}

\author{Yasunobu Uchiyama}
\affil{Department of Physics, Rikkyo University, 3-34-1 Nishi Ikebukuro, Toshima, Tokyo 171-8501, Japan}

\author{Brian J. Williams}
\affil{NASA Goddard Space Flight Center, Code 662, Greenbelt, MD 20771, USA}



\begin{abstract}
We report on {\it NuSTAR} observations of  the mixed morphology supernova remnant (SNR) W49B, focusing on its nonthermal emission. 
Whereas radio observations as well as recent gamma-ray observations evidenced particle acceleration in this SNR, nonthermal X-ray emission 
has not been reported so far.
With the unprecedented sensitivity of {\it NuSTAR} in the hard X-ray band, we detect a significant power-law-like component extending up to 
$\sim 20~{\rm keV}$, most probably of nonthermal origin. 
The newly discovered component has a photon index of $\Gamma =1.4^{+1.0}_{-1.1}$ with an energy flux between 10 and 20~keV of 
$(3.3 \pm 0.7) \times 10^{-13}~{\rm erg}~{\rm cm}^{-2}~{\rm s}^{-1}$. 
The emission mechanism is discussed based on the {\it NuSTAR} data combined with those in other wavelengths in the literature. 
The {\it NuSTAR} data, in terms both of the spectral slope and of the flux, are best interpreted as nonthermal electron bremsstrahlung. 
If this scenario is the case, then the {\it NuSTAR} emission provides a new probe to sub-relativistic particles accelerated in the SNR. 
\end{abstract}


\section{Introduction}
\begin{figure*}[bt]
\epsscale{1.1}
\plotone{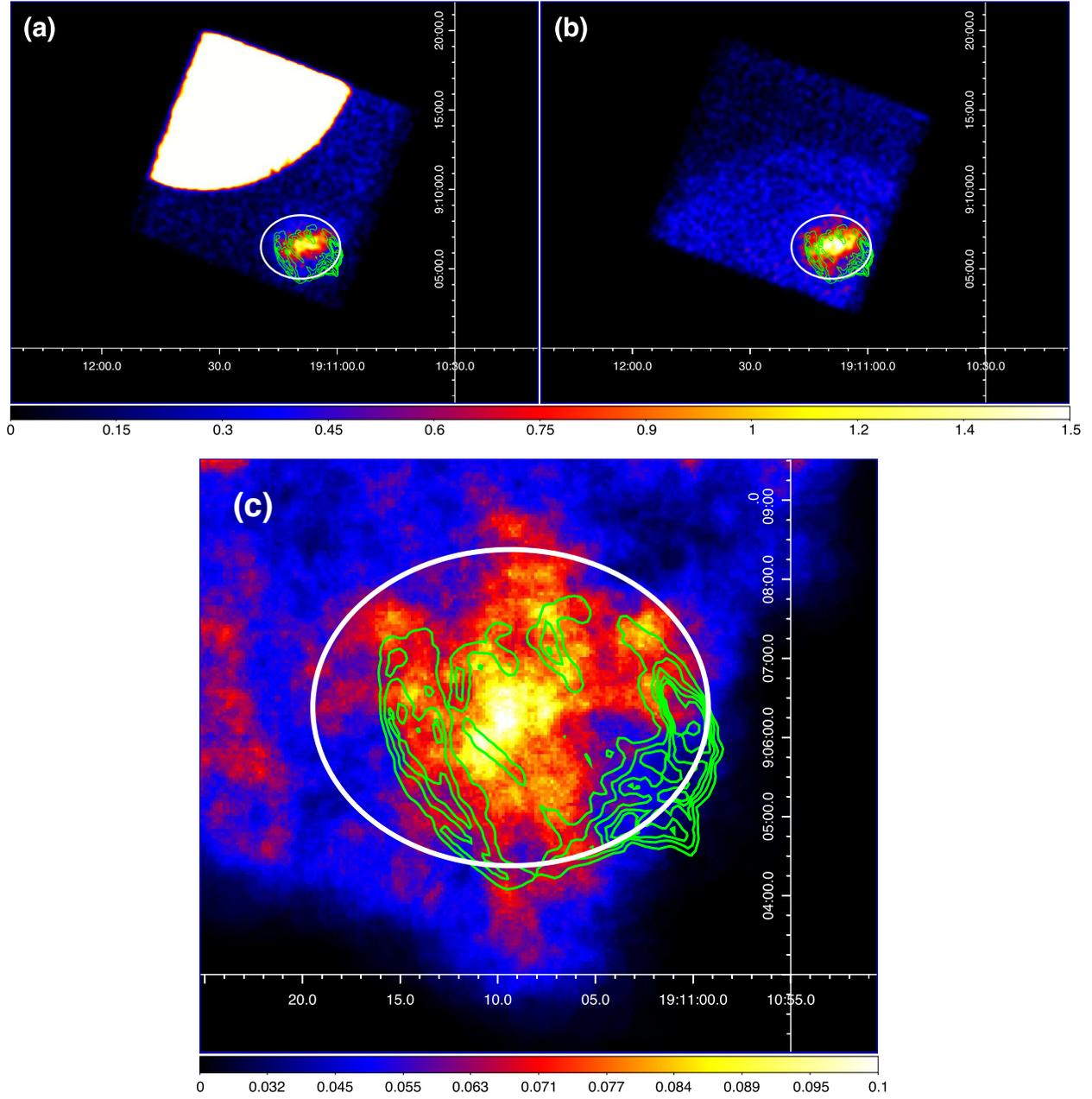}
\caption{(a) Smoothed counts maps in 9--20~keV obtained with FPMA. (b) Same as (a) but obtained with FPMB. (c) Smoothed counts maps (FPMA + FPMB) in 15--20~keV. North is up and the east is to the left. The Gaussian function is used as the smoothing kernel in panels (a) and (b), whereas the top-hat function is used in panel (c). 
The wedge-like feature in panel (a) is due to stray light from GRS~1915+105, while a similar but less bright feature in panel (b) is caused by stray light from 4U~1908+075. 
The contributions from the background except for the stray light component are subtracted from the image in panel (c). The green contours indicate the radio continuum image as observed with the Very Large Array at a frequency of 1.4~GHz in the Multi-Array Galactic Plane Imaging Survey \citep{White2005}. The source extraction region used in the spectral analysis is shown as the white ellipses. 
\label{fig:image}}
\end{figure*}

Particle acceleration in supernova remnants (SNRs) has extensively  been studied with X-ray and 
gamma-ray observations \citep[e.g.,][]{Reynolds2008,Aharonian2013}. 
In the X-ray band, synchrotron radiation has almost exclusively been used as a channel to probe 
electrons accelerated in SNR shocks. 
Accelerated electrons are able to shine also in gamma rays through inverse Compton scattering (IC)
mainly of the cosmic microwave background (CMB) or through bremsstrahlung. 
The hadronic component of accelerated particles can be probed with gamma rays resulting from 
the decay of $\pi^0$ mesons produced by interactions between accelerated protons/nuclei and ambient 
gas as evidenced by the characteristic spectral shape detected with {\it Astro-rivelatore Gamma a Immagini Leggero} ({\it AGILE}) 
Gamma-Ray Imaging Detector \citep{Agile_pion} and 
{\it Fermi Gamma-ray Space Telescope} Large Area Telescope \citep[{\it Fermi} LAT;][]{Fermi_pion}. 

The detection in gamma rays of SNR W49B by the {\it Fermi} LAT \citep{Fermi2010,HESS2018} and by H.E.S.S. \citep{HESS2018} made 
this object an interesting case for studies regarding particle acceleration. 
W49B is one of the most luminous gamma-ray-emitting SNRs in the Galaxy with $L_{\gamma} = 2 \times 10^{35}\, (D/10~{\rm kpc})^2~{\rm erg}~{\rm s}^{-1}$, 
which requires a remarkably large energy density of emitting particles (either electrons or protons) of $U_{e, p} > 10^4~{\rm eV}~{\rm cm}^{-3}$ \citep{Fermi2010}. 
The gamma-ray emission can be interpreted either as $\pi^0$ decay or as electron bremsstrahlung \citep{Fermi2010,HESS2018}. 
Although the spectral break found at 300~MeV is suggestive of the former, the gamma-ray emission mechanism is still not conclusive \citep{HESS2018}. 

Previous X-ray studies of W49B \citep[e.g.,][]{Hwang2000,Miceli2006,Keohane2007,Lopez2013} focused on its bright thermal emission from the shock-heated plasma, 
which was found to be in a recombination-dominant state by \cite{Ozawa2009}. 
Hard X-ray observations at $> 10~{\rm keV}$ are essential in order to search for possible nonthermal radiation for less contamination 
from the thermal emission. 
As discussed by \cite{Uchiyama2002a,Uchiyama2002b}, among a few possible emission mechanisms in the hard X-ray band, 
nonthermal (inverse) bremsstrahlung from sub-relativistic particles is promising in the case of SNRs such as W49B, which is interacting with dense gas \citep{Reach2006,Keohane2007}. 
If detected, the bremsstrahlung component plays a role in disentangling the possible scenarios for the gamma-ray emission and also provides 
a probe to sub-relativistic portion of accelerated particles, which is not accessible with the above mentioned X-ray or gamma-ray emission channels. 

In this Letter, we report on results from recently performed {\it Nuclear Spectroscopic Telescope Array} \citep[{\it NuSTAR};][]{NuSTAR} observations 
of W49B, concentrating on its nonthermal aspect. 
A study of the thermal emission from the same observation is reported in a separate paper by \cite{Yamaguchi2018}. 
Uncertainties quoted in the text and tables, as well as those plotted in figures, indicate 1$\sigma$ confidence intervals.

\section{Observations and Data Reduction}
We performed the {\it NuSTAR} observations of W49B on 2018 March 17--20 (Observation ID: 40301001002; PI: H.~Yamaguchi). 
We reduced the data using the NuSTARDAS v.1.8.0 software package and the calibration database released on 2018 April 19. 
We reprocessed the data with the {\tt nupipeline} tool in the software package. 
We discarded high background periods by applying a filter comparable to the {\tt saamode = optimized} and {\tt tentacle = yes} 
options in {\tt nupipeline}. 
The effective exposure time after the filtering is 122~ks. 

Figure~\ref{fig:image} shows {\it NuSTAR} counts maps. 
A notable feature in the 9--20 keV band is stray light from the high-mass X-ray binary 4U~1908+075, which overlaps with W49B in the focal plane module B (FPMB) data. 
Although focal plane module A (FPMA) data also suffer from stray light from the microquasar GRS~1915+105, 
it does not affect the source extraction region. 
In the 15--20~keV band, where the thermal emission is almost negligible (see \S\ref{sec:ana}), 
a clear excess at the location of the SNR can clearly be seen.

\begin{figure*}[tb]
\epsscale{1.1}
\plottwo{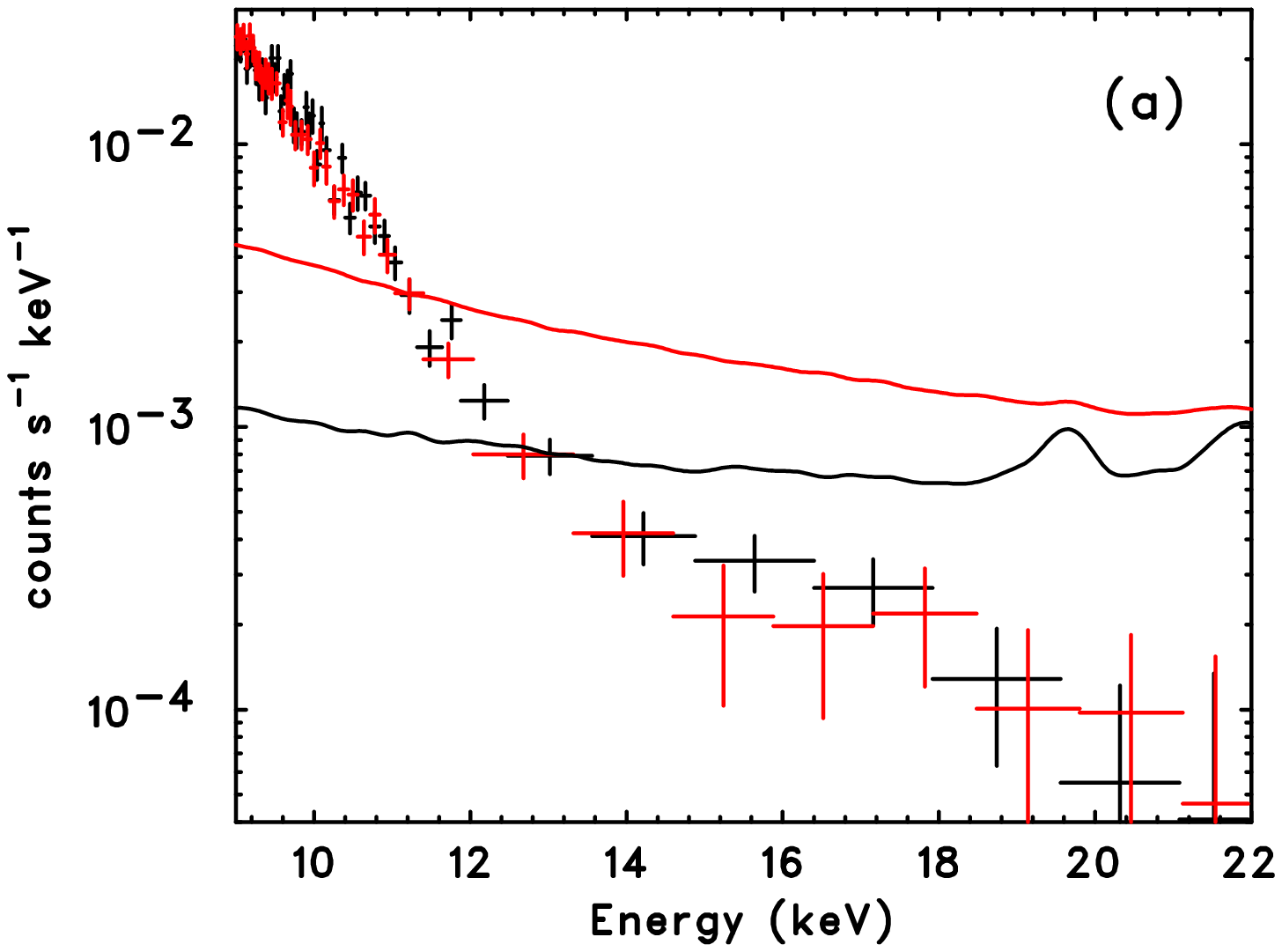}{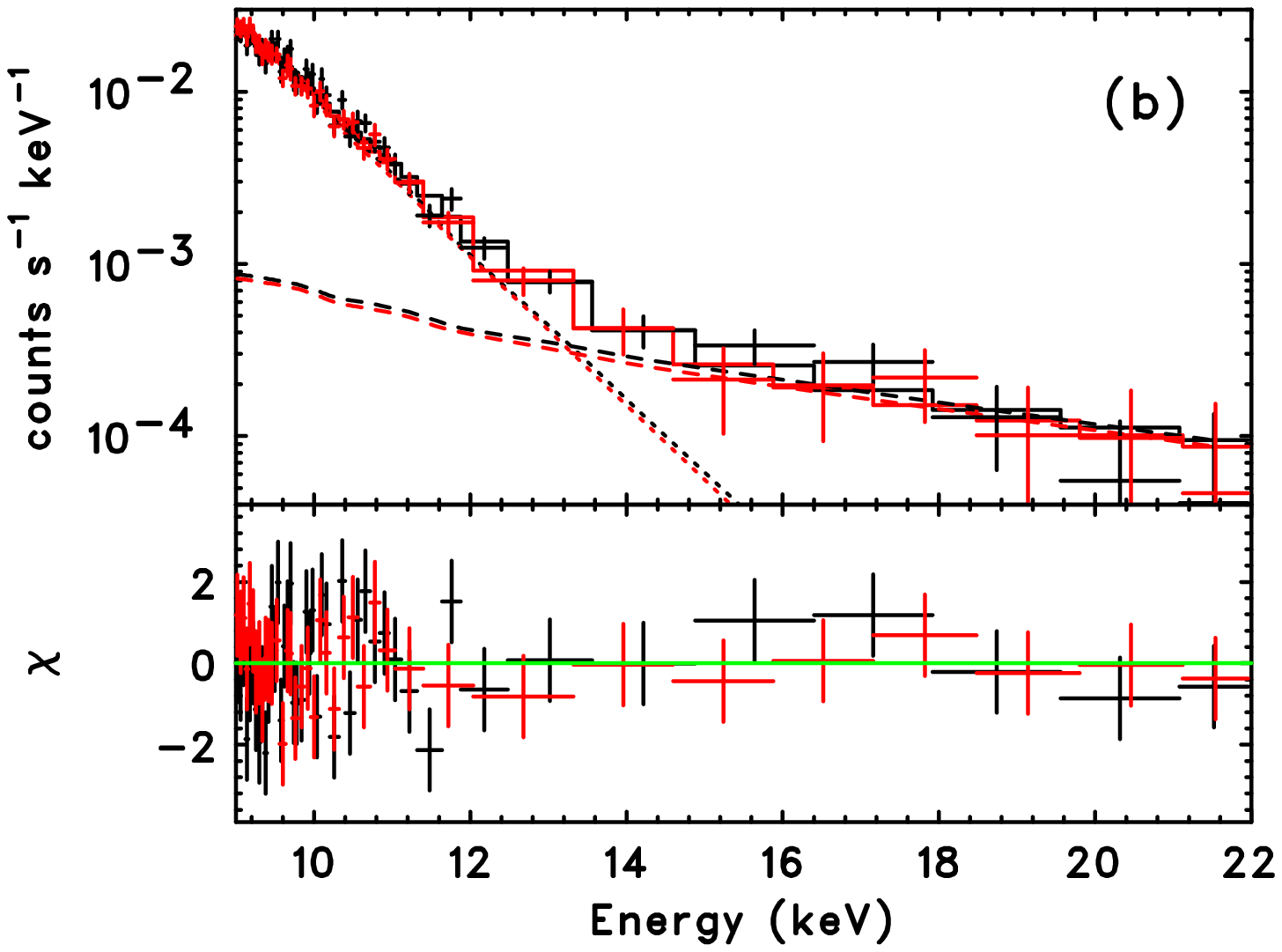}
\caption{(a) Background-subtracted spectra of W49B from {\it NuSTAR} FPMA (black) and FPMB (red). The solid curves are the background models estimated with {\tt nuskybgd}. (b) The same as the left figure but plotted with the best-fit model summarized in Table~\ref{tab:fit_param}. The dotted and dashed curves indicate the RRC and power-law components, respectively. The bottom panel shows residuals from the model. 
\label{fig:spec}}
\end{figure*}

\begin{deluxetable}{ccc}[tbh]
\tablecolumns{3}
\tablewidth{0pt}
\tablecaption{Best-fit parameters\label{tab:fit_param}}
\tablehead{
Component & Parameter & Value
 }
\startdata
RRC &$\varepsilon_{\rm edge}$ & $8.83~{\rm keV}$ (fixed) \\ 
  & $kT_e$\tablenotemark{a}  &  $1.08^{+0.04}_{-0.05}~{\rm keV}$ \\
 & Norm\tablenotemark{b} & $(1.07^{+0.04}_{-0.06})\times10^{-4}~{\rm ph}~{\rm cm}^{-2}~{\rm s}^{-1}$ \\
Power Law & $\Gamma$ &  $1.4^{+1.0}_{-1.1}$ \\
 & Norm\tablenotemark{c} & $(3.3 \pm 0.7) \times 10^{-13}~{\rm erg}~{\rm cm}^{-2}~{\rm s}^{-1}$ \\
 Constant & Factor\tablenotemark{d} & $0.98 \pm 0.03$ \\
 \hline
  & c-stat & 528 \\
  & $\chi^2$ & 582 \\
  & d.o.f. & 605 
\enddata
\tablenotetext{a}{Electron temperature.}
\tablenotetext{b}{Integration from $\varepsilon_{\rm edge}$ to infinity.}
\tablenotetext{c}{Energy flux integrated from 10 to 20~keV.}
\tablenotetext{d}{A constant factor multiplied to the model for the FPMB spectrum to account for possible cross-normalization uncertainties between FPMA and FPMB.}
\end{deluxetable}

\section{Analysis and Results}\label{sec:ana}
Figure~\ref{fig:spec}a shows background-subtracted spectra of W49B obtained with FPMA and FPMB in the energy range between 9 and 22~keV.  
The source extraction region encompasses the whole {\it NuSTAR} emission as indicated in Figure~\ref{fig:image}. 
The background models, plotted in Figure~\ref{fig:spec}a with the data, were estimated using the {\tt nuskybgd} script\footnote{https://github.com/NuSTAR/nuskybgd} \citep{Wik2014}.  
The script provides models consisting of instrumental background, focused X-ray background, and stray light components. 
To model these components, spectra were extracted from three separate regions in each 
telescope focal plane, each square region centered on the detectors not containing W49B. 
The {\tt nuskybgd} software adjusts the normalization of each standard background component 
based on fits to these spectra. In order to account for the stray light, additional spectral models, 
with appropriate responses, were manually added for the emission of the two stray light sources 
so that the background solution would not be biased by their extra flux.  The best-fit model parameters 
for 4U 1908+075 in the FPMB observation were then used to add its contribution to the background 
spectra for all FPMB extraction regions of W49B by scaling it by the area of the region. 
This is because the stray light, being undeflected, produces a uniform pattern in the focal plane. 
The higher background of FPMB than that of FPMA in Figure~\ref{fig:image} is attributed to the stray light contaminating the source extraction region. 

Both FPMA and FPMB fluxes at $\gtrsim 15~{\rm keV}$ appear to be higher than the extrapolations of the lower-energy data points, 
suggesting the presence of a hard tail in addition to the thermal component. 
To validate our background modeling, we extracted spectra from a region outside W49B, and also estimated the background for the 
region by running {\tt nuskybgd}. 
We found that the spectra are consistent with the background model and that no significant hard tail emission is detected in the region, confirming 
the accuracy of the background model. 
Another demonstration of the background model accuracy comes from the consistency of the results from FPMA and FPMB (Figure~\ref{fig:spec}) 
in spite of the different levels of the stray light contamination between the two sensors.

We fitted the spectra with a model composed of a thermal component and a power law. 
Following the recommendation found in the {\it NuSTAR} Analysis webpage,\footnote{https://heasarc.gsfc.nasa.gov/docs/nustar/analysis/} 
we multiplied a constant factor to the model for the FPMB data, and allowed it to vary in order to account for possible cross-normalization uncertainties. 
As the thermal component, we employed a recombination edge emission model, 
{\tt redge} in the {\tt XSPEC} package \citep{Arnoud1996},  
considering the result by \cite{Ozawa2009} that the radiative recombination continuum (RRC) of He-like Fe is the dominant 
thermal component in this energy range. 
The edge energy of the RRC ($\varepsilon_{\rm edge}$) was fixed at 8.83~keV. 
In the fitting procedure whose results are presented below, 
we included the background components predicted by {\tt nuskybgd} as a model rather than subtracting it, 
and performed a maximum likelihood fitting based on the Cash statistic \citep{Cash1979}. 
Before fitting, the background models were smoothed with the ``353QH twice'' algorithm \citep{Friedman1974} 
in order to remove artificial small structures due to statistical fluctuations.  
As a cross check, we also fitted the background-subtracted spectra using the W statistic \citep{Wachter1979}, 
in which the background in each energy bin is supposed to be expressed with its own parameter.
We confirmed that the two results are consistent with each other. 

The best-fit models are overlaid on the background-subtracted spectra in Figure~\ref{fig:spec}b and the best-fit parameters are 
summarized in Table~\ref{tab:fit_param}. 
The fit gave a relatively hard photon index ($\Gamma$) of the power-law component despite the large statistical error. 
A fit without a power law (null hypothesis) was also performed, yielding a c-stat of $C_0 = 580$ as compared to a c-stat of $C_1 = 528$ from the fit with 
a power law (alternative hypothesis). 
Thus, the test statistic (TS) of the power-law component is ${\rm TS} = C_0 - C_1 = 52$. 
In order to quantify the statistical significance of the power-law component, we ran Monte-Carlo simulations and 
generated $10^4$ spectra assuming the null hypothesis. 
We fit each of the simulated spectra with the models for the null and alternative hypotheses, and 
calculated TS in the same manner as for the observational data. 
We found TS only up to 13 in the simulated datasets, which indicates that the null hypothesis probability is less than $10^{-4}$.

\section{Discussion}
In the previous section, we described the detection of a hard tail in W49B with {\it NuSTAR}. 
If we interpret the hard tail as a thermal bremsstrahlung emission using the {\tt bremss} model in the {\tt XSPEC} package, we 
obtained 9.0~keV as a lower limit to the electron temperature, which is unrealistically high for an SNR. 
Thus, the emission detected with with {\it NuSTAR} is most likely of nonthermal origin. 

Synchrotron X-rays have been detected in a number of young SNRs, and thus the hard tail emission could be synchrotron radiation. 
Under the assumption that synchrotron cooling is dominant, \cite{Zira2007} gave a synchrotron cutoff energy as 
\begin{eqnarray}\label{eq:shockv}
\varepsilon_0 = 0.55 \left(\frac{V_{\rm s}}{3000~{\rm km}~{\rm s}^{-1}} \right)^2 \eta^{-1}~{\rm keV}, 
\end{eqnarray}
where $V_{\rm s}$ and $\eta\ (\geq 1)$ are the shock speed and the so-called ``gyrofactor'', respectively. 
\cite{Keohane2007} estimated the forward shock velocity in the X-ray emitting plasma to be $V_{\rm s} \sim 1000~{\rm km}~{\rm s}^{-1}$. 
As \cite{Keohane2007} and \cite{Zhu2014} pointed out, the shock velocity should be much slower in the denser regions where infrared lines 
such as [\ion{Fe}{2}] are detected. 
Thus, the above value can be regarded as an upper limit to the shock velocity of this SNR. 
Substituting $V_{\rm s} = 1000~{\rm km}~{\rm s}^{-1}$ in Equation~(\ref{eq:shockv}), we obtain a cutoff energy of $\varepsilon_0  \leq 0.06~{\rm keV}$, 
which is about 
two orders of magnitude lower than the {\it NuSTAR } bandpass. 
The analytical formula for the synchrotron spectrum by \cite{Zira2007}, with the above cutoff energy, predicts that 
synchrotron emission has a steep spectrum corresponding to $\Gamma \sim 5$ at 10~keV, which contradicts the hard {\it NuSTAR} spectrum. 
We, therefore, conclude that synchrotron is an unlikely explanation for the hard tail emission.

\begin{figure*}[tbh]
\epsscale{1.1}
\plottwo{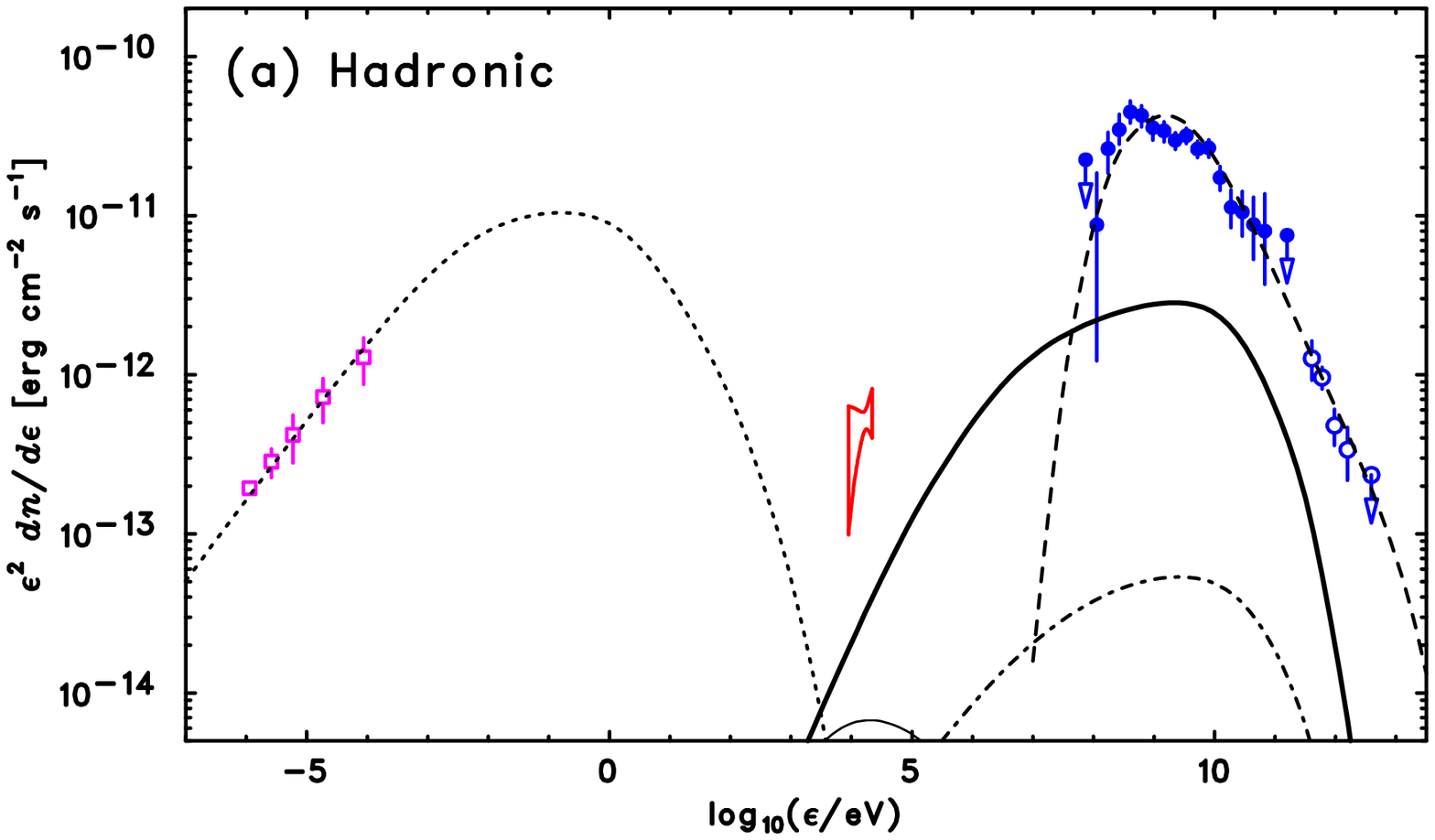}{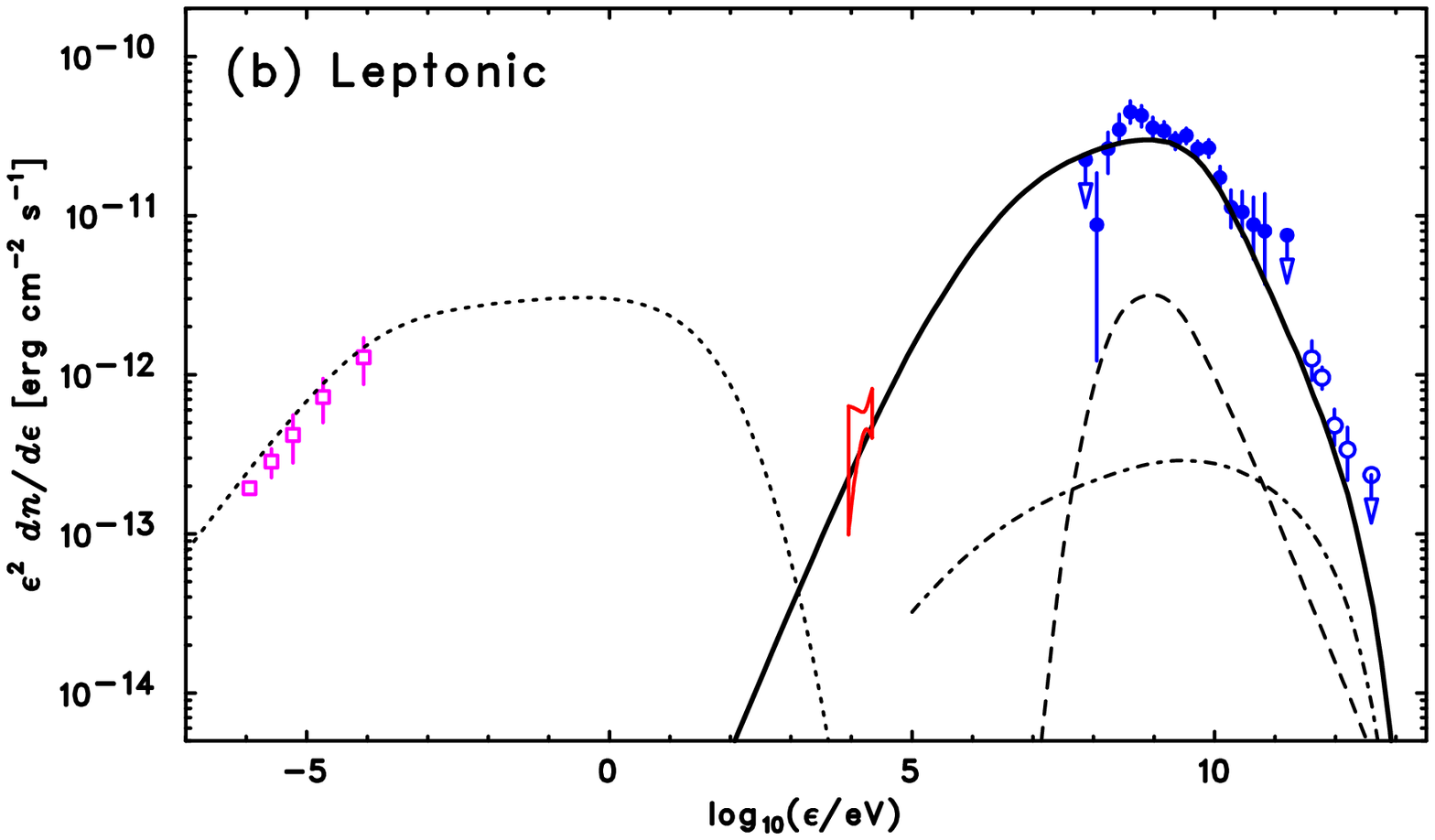}
\caption{Broadband SED of the nonthermal emission from W49B with (a) the hadronic model and (b) the leptonic model. 
The region enclosed by the red curves corresponds to a 68\% confidence region of the spectral parameters calculated from the covariance matrix. 
The radio data in magenta are taken from \cite{Moffett1994}. The blue points are gamma-ray data from {\it Fermi} LAT (filled circles) and H.E.S.S. (open circles) 
reported by \cite{HESS2018}. The black curves represent each component of the emission models: electron bremsstrahlung (thick solid line), proton bremsstrahlung (thin solid line), synchrotron (dotted line), IC (dotted-dashed line), and $\pi^0$ decay (dashed line).  
\label{fig:sed}}
\end{figure*}

\begin{deluxetable*}{cccccccccc}[tbh]
\tablecolumns{9}
\tablewidth{0pt}
\tablecaption{Parameters for the Models \label{tab:model_param}}
\tablehead{
Model & $s_1$ & $s_2$ & $p_{\rm b}$ & $p_{0e}$ & $p_{0p}$ & $B$ & $n$\tablenotemark{a} & $W_p$\tablenotemark{b} &$K_{ep}$ \\
   &     &   &  (${\rm GeV}~c^{-1}$) &  (${\rm TeV}~c^{-1}$) & (${\rm TeV}~c^{-1}$) & ($\mu{\rm G}$) & (${\rm cm}^{-3}) $ & ($10^{49}~{\rm erg}$) &  
 }
\startdata
Hadronic & 2.0 & 2.8 & 30 & 5 & 100 & 100 & $ 100 $ & $ 12 $ & 0.01\\
Leptonic & 2.0 & 2.9 & 10 & 10 & 100 & 25 & 100 &  $ 0.94 $ & 1.0 
\enddata
\tablenotetext{a}{Ambient gas density.}
\tablenotetext{b}{Total kinetic energy of radiating protons integrated above 10~MeV. The distance to W49B is assumed to be 10~kpc.}
\end{deluxetable*}

IC would be another possible radiation channel to account for the hard tail emission. 
The photon index of $\Gamma = 1.4^{+1.0}_{-1.1}$ is consistent with the radio index of $\alpha = -0.5$ \citep{Moffett1994} so that 
the spectral slope of the hard X-ray component can be explained by IC from the same electron population as that emitting synchrotron photons in the radio band. 
However, this scenario faces difficulty in terms of energetics. 
When we consider the CMB and the interstellar radiation field \citep[ISRF; e.g.,][]{Porter2006} as the seed photons, 
the radiating electrons are required to have a huge total energy of $> 10^{51}~{\rm erg}$ in order to raise the IC flux to the level of the hard X-ray flux we observed. 
Thus, the IC scenario is also unlikely. 

The dense gas environment around W49B makes nonthermal bremsstrahlung either by electrons or protons 
a viable option as the scenario for the hard tail emission. 
Indeed, the hard spectrum is consistent with this scenario. 
If particles have a power-law spectrum in the form of $dN/dE \propto E^{-s}$, their bremsstrahlung 
spectrum becomes also a power law ($dn/d\varepsilon \propto \varepsilon^{-\Gamma}$) with $\Gamma \sim s$. 
Assuming the canonical value for the the spectral index of the particle energy distribution from diffusive shock acceleration, 
$s \simeq 2$, we expect the bremsstrahlung spectrum has $\Gamma \simeq 2$. 
However, because of the ionization loss, the particle spectrum is ``loss-flattened'' below a break energy 
$E_{\rm br}$, which is determined by equating the ionization loss timescale and particle injection timescale \citep{Uchiyama2002a,Uchiyama2002b}. 
Therefore, bremsstrahlung spectra below the corresponding break should be hard with $\Gamma \sim 1$. 

For more quantitative discussion, we calculate emission models to explain the spectral energy distribution (SED) of nonthermal 
radiation of W49B from radio to gamma rays. 
Protons and electrons are injected to the emitting region with a constant luminosity. 
We assumed the injection spectra in the form of 
\begin{eqnarray}\label{eq:kinetic}
Q_{e,p} = A_{e,p} \left( \frac{p}{1~{\rm GeV}~c^{-1}}\right)^{-s_1} \left[ 1 + \left(\frac{p}{p_{\rm{b}}}\right)^2 \right]^{(s_1 - s_2)/2} \exp\left( - \frac{p}{p_{0e,p}} \right), 
\end{eqnarray}
which has a smooth spectral break at $p_{\rm b}$ and an exponential cutoff at $p_0$. 
We define the electron-to-proton ratio as $K_{ep} \equiv A_e/A_p$. 
The particle spectra are deformed as a result of radiative and nonradiative cooling. 
The kinetic equations for protons and electrons, 
\begin{equation}
\frac{\partial N_{e,p}(p, t)}{\partial t} = \frac{\partial}{\partial p} [b_{e,p}(p)\ N_{e,p}(p, t)] + Q_{e,p}(p), 
\end{equation}
where $b_{e,p}$ denotes momentum loss rate, are solved to obtain $N_{e,p} (p, t)$, particle spectra after the deformation.  
We take into account cooling by ionization, bremsstrahlung, synchrotron, IC, and $\pi^0$ decay to calculate $b_{e,p}$. 
We solve Equation~(\ref{eq:kinetic}) for $t = 2000~{\rm years}$ to obtain $N_{e,p} (p, t)$, and  we then calculate radiation spectra 
of bremsstrahlung, synchrotron, IC, and $\pi^0$ decay. 
The prescriptions by \cite{Kamae2006} are used for the calculation of $\pi^0$-decay spectra. 
In addition to the CMB, we include the ISRF at the location of W49B taken from GALPROP \citep{Porter2006} as seed photons for IC.

We present the calculation results overlaid on multi-wavelength data including the {\it NuSTAR} data in Figure~\ref{fig:sed}, and 
summarize the model parameters in Table~\ref{tab:model_param}. 
We here show two models: the gamma-ray emission is predominantly ascribed to $\pi^0$ decay in one model (hadronic model; Figure~\ref{fig:sed}a)  
and to electron bremsstrahlung in the other model (leptonic model; Figure~\ref{fig:sed}b). 
In the calculation, we assumed the gas density of $n = 100~{\rm cm}^{-3}$, which is roughly consistent with the estimate by \cite{HESS2018}. 
The magnetic field strength was determined so that the synchrotron flux match the radio data. 
The other parameters concerning the particle spectra were chosen so that the shapes of the model curves match the data. 
Although the southwestern corner of W49B is close to the field of view of {\it NuSTAR}, we assumed that the entire emission of the SNR is covered. 
This assumption would be justified to some extent as the best-fit positions of the gamma-ray emissions, including that of the {\it Fermi} LAT, which is 
located near the western edge of the SNR, are within the field of view. 
Both models reproduce the spectral slope of electron bremsstrahlung in the hard X-ray band consistent with the {\it NuSTAR} measurement. 
However, the two models predict largely different fluxes in the {\it NuSTAR} bandpass. 
The leptonic model nicely fits the {\it NuSTAR} data, whereas the electron bremsstrahlung component of the hadronic model falls short of 
the observed hard X-ray flux by about one order of magnitude. 
The contribution from proton bremsstrahlung is almost negligible even in the hadronic model. 
We conclude that the leptonic model fits better the data as far as a simple one-zone model is considered.

The leptonic model plotted in Figure~\ref{fig:sed} (b) requires a large electron-to-proton ratio of $K_{ep} \sim 1$. 
This challenges the current understanding of diffusive shock acceleration as electrons are generally difficult to inject into an acceleration process \citep[e.g.,][]{Park2015}. 
In order to avoid this and to make the hadronic model a possible option, 
another electron population that is accelerated only up to sub-relativistic energies would be helpful. 
If we assume a cutoff at $\sim {\rm MeV}$ in the spectrum of the second electron population, the electrons 
shine only in the hard X-ray band through bremsstrahlung with negligible contributions to the radio band through synchrotron 
and to the gamma-ray band through bremsstrahlung. 
We emphasize that, even in this case, electron bremsstrahlung is the most plausible emission process to account for the hard X-ray data.

If electron bremsstrahlung is indeed the emission mechanism responsible for the {\it NuSTAR} emission, the radiating electron population
should be in the sub-relativistic regime with kinetic energies of $\sim 10~{\rm keV}$. 
In addition to bremsstrahlung, those electrons can cause K-shell ionization of ambient Fe atoms and can emit the K$\alpha$ line at 6.4~keV 
\citep{Dogiel2011,Nobukawa2018,Okon2018,Saji2018}. 
Nonthermal bremsstrahlung in the hard X-ray band, therefore, should always be accompanied by the neutral Fe K$\alpha$ line. 
The equivalent width (EW) of the line with respect to the nonthermal bremsstrahlung component is ${\rm EW} \lesssim 400~{\rm eV}$ if the Fe abundance 
is consistent with solar \citep{Dogiel2011}. 
By extrapolating the power law to lower energies and assuming an emission line at 6.4~keV with ${\rm EW}  = 400~{\rm eV}$, 
we found that {\it NuSTAR} cannot detect the line because of the bright thermal emission. 
Analyzing the {\it NuSTAR} data below 9~keV, we indeed did not see any hints of a line structure at 6.4~keV. 
It is of interest to search for the line in data taken with charge-coupled device cameras on board other operating observatories, which have better energy resolution at that energy. 
Eventually, X-ray micro-calorimeters on board future X-ray astronomy satellites such as {\it XRISM} (formerly known as {\it XARM}) and {\it Athena} can easily detect the line. 
Detection of the neutral Fe K$\alpha$ line as well as its EW with respect to the continuum detected by {\it NuSTAR} will help us confirm the radiation mechanism 
of the hard X-ray emission and will provide us with further information about the spectra of sub-relativistic particles accelerated in this SNR.

\acknowledgments
We appreciate the {\it NuSTAR} SOC members for their support. We are grateful Fran\c{c}ois Brun for providing us with the {\it Fermi} LAT and H.E.S.S. data points used in this Letter. We thank Shigeo Yamauchi, Masayoshi Nobukawa, and Katsuji Koyama for proposal preparation. We also thank Shiu-Hang Lee for useful discussions.



\end{document}